
\magnification=1200
\baselineskip=20pt
\def\sqr#1#2{{\vcenter{\vbox{\hrule height.#2pt
        \hbox{\vrule width.#2pt height#1pt \kern#1pt
          \vrule width.#2pt}
        \hrule height.#2pt}}}}

\def\boxit#1{\vbox{\hrule\hbox{\vrule\kern3pt
   \vbox{\kern3pt#1\kern3pt}\kern3pt\vrule}\hrule}}
\def\lsim{<\kern-2.5ex\lower0.85ex\hbox{$\sim$}\ }
\def\rsim{>\kern-2.5ex\lower0.85ex\hbox{$\sim$}\ }
\overfullrule=0pt
\def\LAMBDABAR {\hbox{$\lambda$\kern-0.52em\raise+0.45ex\hbox{--}\kern+0.2em}}
\def\partialbar {\hbox{$\partial$\kern-0.52em\raise0.2ex\hbox{/}\kern+0.1em}}
\def\abar {\hbox{A\kern-0.6em\raise0.2ex\hbox{/}\kern+0.1em}}
\def\ebar {\hbox{E\kern-0.6em\raise0.2ex\hbox{/}\kern+0.1em}}
\centerline{\bf POSSIBLE TESTS OF CP-VIOLATIONS IN $B^\pm$ DECAYS}

\vskip 1.5cm
\centerline{Susumu Okubo}
\centerline{Department of Physics and Astronomy}
\centerline{University of Rochester}
\centerline{Rochester, NY 14627, USA}
\vskip 3cm
\centerline{\bf Abstract}

Some experimental tests are suggested for possible CP-violations for decays
of $B^\pm$ bosons. Especially, a detailed comparison of decays
$B^+ \rightarrow D^0 \pi^+ \pi^+ \pi^-$ and $B^- \rightarrow
\overline{D^0} \pi^- \pi^- \pi^+$ will lead to tests of not only CP, but
also C, and P invariances.

\vfill\eject

One pressing problem facing the so-called $B$-physics is the question of
the CP-violation.  The most likely and most interesting place to find it is
in the neutral sector of $B^0 - \overline{B^0}$
complex just as in the familiar $K^0 - \overline{K^0}$ complex.  However,
we will consider, in this note, possible ways to detect CP-violations in
the charged sector $B^+$ and its anti-particle $\overline{B^+} (\equiv
B^-)$.  For this end, it may be useful to refresh our mind of known facts
about the corresponding problems in $K^\pm$ physics.

First, the TCP invariance alone guarantees$^{1),2)}$ the equivalence of
total
decay rates for a particle and its anti-particle, provided that they are
not neutral.  For example, we will have
$$\eqalignno{&R ( K^+ \rightarrow \ {\rm all})\  = R ( K^- \rightarrow
\ {\rm all}) &(1)\cr
&R (\Sigma^+ \rightarrow \ {\rm all} )\  =
R( \overline{\Sigma^+} \rightarrow \ {\rm all} ) &(1^\prime)\cr}$$
for total
 decay rates.  However, we also know the following results$^{2),3),4)}$:

\item{(i)} Individual exclusive decay rates may differ$^{2)}$
 in general from their anti-particle ones, i.e.
$$\eqalignno{&R(K^+ \rightarrow \pi^+ \pi^+ \pi^- ) \not= R(K^-
\rightarrow \pi^- \pi^- \pi^+) &(2)\cr
&R(\Sigma^+ \rightarrow p \pi^0 ) \not= R(\overline{\Sigma^+}
\rightarrow \overline p  \pi^0) &(3)\cr}$$

\item{  }unless  C or CP   is conserved in the process.  However we have
$$R(K^+ \rightarrow \pi^+ \pi^0) = R (K^- \rightarrow \pi^- \pi^0)
\eqno(4)$$

\item{  } by the TCP alone, if we ignore the electromagnetic interaction.  The
reason is that the $2 \pi \leftrightarrow 3 \pi$ process is forbidden by the
 $G$-parity.$^{2)}$

\item{(ii)} The Dalitz plot for $K^+ \rightarrow \pi^+ \pi^+ \pi^-$ decay
will be different$^{3)}$ from that for $K^- \rightarrow \pi^- \pi^- \pi^+$
 under the same assumption.

\item{(iii)} For particles with non-zero spins, asymmetry parameters for a
particle and its anti-particle are in general different, unless the CP
 or C is
again conserved.  For example, consider the decay $\Sigma^+ \rightarrow p
\pi^0$ and its anti-particle mode $\overline{\Sigma^+} \rightarrow
\overline p \pi^0$.  Writing their decay matrix elements as
$$\eqalignno{&M (\Sigma^+ \rightarrow p \pi^0) = A + B \  \underline{\sigma}
\cdot \underline{k} \quad , &(5)\cr
&M (\overline{\Sigma^+} \rightarrow \overline p
 \pi^0) = \overline A + \overline B \ \underline{\sigma}
\cdot \underline{k} \quad , &(5^\prime)\cr}$$
we will have$^{2)}$
$$\overline A = -A \quad , \quad \overline B = B \eqno(6)$$
only if the CP-conservation holds valid.  Here, $\underline{k}$ is the pion
momentum in the rest frames of $\Sigma^+$ and $\overline{\Sigma^+}$.
Defining the asymmetry parameters $\alpha$ and $\overline \alpha$ by
$$\alpha = {2 {\rm Re} (AB^*) \over
| A |^2 + |B|^2}\quad ,
\quad \overline \alpha = {2 {\rm Re} (\overline A\  \overline B^*)
\over |\overline A |^2 + | \overline B |^2} \eqno(7)$$
we will have$^{4)}$
$$\overline \alpha  = - \alpha \quad , \eqno(8)$$
if the CP invariance holds.  Note that the C-invariance
predicts $\overline \alpha = \alpha$, in constrast.

These facts can be used in principle to detect possible CP-violations.
However, the degree of the violations in the $K$-sector is very small by a
factor of $10^{-3}$ in comparison to the CP-conserving one so that they are
in reality too difficult for experimental tests.  On the other sides the
CP-violations for the $B$-sectors may be considerably larger and it may be
possible to experimentally detect them.  This is the purpose of this note.

The possible proposed decay modes for detection of CP violations in the
$B$-physics are as follows:

\item{(i)} Comparison of decay rates for $B^+ \rightarrow D^0 \pi^+$ and
$B^- \rightarrow \overline{D^0} \pi^+$ as well as for $B^+ \rightarrow D^+
\pi^+ \pi^-$ and $B^- \rightarrow D^- \pi^- \pi^+$ as in
Eqs. (2) and (3).

\item{(ii)} Comparison of Dalitz plots for $B^+ \rightarrow D^+ \pi^+ \pi^-
$ and $B^- \rightarrow D^- \pi^- \pi^+$ just as in the $K^\pm \rightarrow 3
\pi$ decays. In case of inequivalence, then both C and CP are violated.

\item{(iii)} Comparison of polarization correlations for
$$\eqalignno{&B^+ \rightarrow (D^* )^+ \rho^0 \rightarrow (D^0 \pi^+)(\pi^+
\pi^- ) \quad , &(9)\cr
&B^- \rightarrow \overline{(D^* )^+} \rho^0 \rightarrow
 (\overline{D^0} \pi^-) (\pi^- \pi^+ ) \quad . &(9^\prime)\cr}$$
Since (i) and (ii) above are essentially the same as in the cases of
$K$-decays, we will explain here only (iii).  The decay matrix elements for
$B^+ \rightarrow (D^*)^+ \rho^0$ is expressed as
$$M \left( B^+ \rightarrow (D^*)^+ \rho^0 \right) =
A (\underline{\varepsilon_1} \cdot \underline{\varepsilon_2}) +
B (\underline{\varepsilon_1} \cdot \underline{k} ) (
\underline{\varepsilon_2} \cdot \underline{k}) +
C (\underline{\varepsilon_1} \times \underline{\varepsilon_2} )
 \cdot \underline{k} \eqno(10)$$
for some constants $A,\ B, \ {\rm and}\ C$,
where $\underline{\varepsilon_1}$ and $\underline{\varepsilon_2}$ are
polarization vectors of $(D^*)^+$ and $\rho^0$, respectively, and
$\underline{k}$ is the $\rho^0$-momentum in the rest frame of
$B^+$.  The corresponding decay matrix element for $B^-$ is also written as
$$M \left( B^- \rightarrow \overline{(D^*)^+} \ \rho^0 \right) =
\overline A (\underline{\varepsilon_1} \cdot \underline{\varepsilon_2} )
+ \overline B  (\underline{\varepsilon_1} \cdot \underline{k} )
 (\underline{\varepsilon_2} \cdot \underline{k} ) +
 \overline C
 (\underline{\varepsilon_1} \times \underline{\varepsilon_2} )
 \cdot\underline{k} \quad .\eqno(11)$$
If the CP is conserved, then we will have
$$\overline A = -A \quad , \quad \overline B = -B \quad , \quad
\overline C = C \quad . \eqno(12)$$
Now, we also  write
$$\eqalignno{&M ((D^*)^+ \rightarrow D^0 \pi^+ ) = \alpha
\ \underline{\varepsilon_1} \cdot \underline{p} &(13)\cr
&M (\rho^0 \rightarrow \pi^+ \pi^- ) = \beta \ \underline{\varepsilon_2}
 \cdot \underline{q} &(14)\cr}$$
for some constants $\alpha$ and $\beta$, where $\underline{p}$
 and $\underline{q}$ are momenta of pions, measured in the rest frame of
$(D^*)^+$ and $\rho^0$, respectively.  Since $(D^*)^+ \rightarrow D^0
\pi^+$ is the CP-conserving strong decay, the same formula with the same
coefficients $\alpha$ and $\beta$ holds valid also for the corresponding
anti-particle.  Then, the combined decay matrix elements for $B^\pm
\rightarrow (D^*)^\pm \rho^0$ followed by $(D^*)^+ \rightarrow
D^0 \pi^+$ and $\rho^0 \rightarrow \pi^+ \pi^-$ will be given by
$$M \left( B^+ \rightarrow (D^0 \pi^+)(\pi^+ \pi^-) \right)
= \alpha \beta \left\{ A (\underline{p} \cdot \underline{q}) +
 B (\underline{p} \cdot \underline{k})  (\underline{q} \cdot \underline{k}) +
C  (\underline{p} \times \underline{q})  \cdot \underline{k}
\right\} \eqno(15)$$
$$M \left( B^- \rightarrow (\overline{D^0} \pi^-)(\pi^- \pi^+) \right)
= \alpha \beta \left\{ \overline A (\underline{p} \cdot \underline{q}) +
 \overline B (\underline{p} \cdot
\underline{k})  (\underline{q} \cdot \underline{k})
+  \overline C  (\underline{p} \times \underline{q}) \cdot \underline{k}
\right\} \quad . \eqno(15^\prime)$$
\noindent From these relations, we can test in principle the relation Eq. (12).

\item{(iv)} We can relax the condition of the two-step decay processes
 as follows.  Consider the 4-body decay $B^+ \rightarrow D^0
\pi^+ \pi^+ \pi^-$ and its anti-particle one $B^- \rightarrow
\overline{D^0} \pi^-
\pi^- \pi^+$.  Let $\underline{k_1},\ \underline{k_2},$ and
$\underline{k_3}$ be 3 pion momenta.  Then, the decay matrix elements are
written as
$$\eqalignno{&M(B^+ \rightarrow D^0 \pi^+ \pi^+ \pi^-) = A_0 +
C_0 (\underline{k_1} \times \underline{k_2}) \cdot \underline{k_3}
\quad , &(16)\cr
&M(B^- \rightarrow \overline{D^0} \pi^- \pi^- \pi^+) = \overline{A_0} +
\overline{C_0} (\underline{k_1} \times \underline{k_2}) \cdot \underline{k_3}
\quad . &(16^\prime)\cr}$$
\item{  } Now, $A_0,\ C_0,\ \overline{A_0},\ {\rm and}\ \overline{C_0}$
are functions of
the products $\underline{k_i} \cdot \underline{k_j}\ (i,j=1,2,3)$.
 The Bose statistics of the pions requires that $A_0$ and $\overline{A_0}$
 are symmetric for $\underline{k_1} \leftrightarrow \underline{k_2}$,
while $C_0$ and $\overline{C_0}$ are anti-symmetric for
$\underline{k_1} \leftrightarrow \underline{k_2}$. Here,
$\underline{k_1},\ {\rm and}\ \underline{k_2}$ are momenta of
likely-charged pions.
   If the CP
 is conserved, we will have
then
$$\overline{A_0} = - A_0 \quad , \quad \overline{C_0} = C_0
 \eqno(17)$$
\item{  } We note that the $C$-invariance implies $\overline{A_0}
 = A_0$ and $\overline{C_0} = C_0$ in contrast, while the $P$-invariance
requires $A_0 = \overline{A_0} = 0$. Hence, we may also test C and P
 invariances, separately.  The same remark also applies to analysis
of $D^+ \rightarrow K^0 \pi^+ \pi^+ \pi^-$ and
$D^- \rightarrow \overline{K^0} \pi^- \pi^- \pi^+$ decays.

\item{(v)} Another possible test will be to measure
$$e \ \overline e \rightarrow \Sigma_B +
\overline \Sigma_B \eqno(18)$$
\item{  }followed by
 $\Sigma_B \rightarrow \pi \Sigma_C$ and $\overline \Sigma_B
\rightarrow \pi \overline  \Sigma_C $.  Here, $\Sigma_B$ and
 $\Sigma_C$ are fermions of forms $buu$ and $cuu$, respectively in terms of
quarks.  Analogously to the process (iii), we can measure
 in principle the
CP-violating polarization correlation in this decay.  However
since the production rate for Eq. (18) is expected to be small, it may not
be experimentally feasible and we will not go into detail of the analysis.

\medskip

\noindent {\bf Acknowledgement}

\smallskip

This work is supported in part by the U.S. Department of Energy grant no.

\noindent DE-FG02-91ER40685.

\vfill\eject

\noindent {\bf References}

\smallskip

\item{1.} G. L\"uders and B. Zumino, Phys. Rev. {\bf 106}, (1957) 385.
 T. D. Lee, R. Oehme and C.~N. Yang, ibid {\bf 106}, (1957) 340.

\item{2.} S. Okubo, Phys. Rev. {\bf 109}, (1958) 984.

\item{3.} Y. Ueda and S. Okubo, Phys. Rev. {\bf 139}, (1965) B1591.

\item{4.} C. K. Chao, Nucl. Phys. {\bf 9}, (1958/59) 652.

\end